\documentclass[prd,showpacs]{revtex4}
\usepackage{amsmath,amssymb}
\usepackage{xcolor}

\numberwithin{equation}{section}
\begin{document}
\title{Magnetic corrections to the fermionic Casimir effect in Horava-Lifshitz theories}
\author{Andrea Erdas}
\email{aerdas@loyola.edu}
\affiliation{Department of Physics, Loyola University Maryland, 4501 North Charles Street,
Baltimore, Maryland 21210, USA}
%\date{May, 2023}
\begin {abstract} 
In this paper I investigate the effect of a magnetic field on the Casimir effect due to a massless and charged fermion field that violates Lorentz invariance according to the Horava-Lifshitz theory. I focus on the case of a fermion field that obeys MIT bag boundary conditions on a pair of parallel plates. I carry out this investigation using the $\zeta$-function technique that allows me to obtain Casimir energy and pressure in the presence of a uniform magnetic field orthogonal to the plates. I investigate the cases of the parameter associated with the violation of Lorentz invariance being even or odd and the cases of weak and strong magnetic field, examining all possible combinations of the above quantities.
In all cases I obtain simple and very accurate analytic expressions of the magnetic field dependent Casimir energy and pressure.
\end {abstract}
\maketitle
%%%%%%%%%%%%%%%%%%%%%%%%%%%%%%%%%%%%%%%%%%%%%%%%%%%%%%%%%%%%%
\section{ Introduction}
\label{1}
The Casimir effect was theoretically predicted 75 years ago \cite{Casimir:1948dh}. In its simplest form, quantum field theory effects produce an attraction between two uncharged conducting plates facing each other in vacuum. The first experimental tests of it \cite{Sparnaay:1958wg} were done ten years after Casimir's theoretical work and were relatively consistent with the theoretical prediction. The effect has been successfully confirmed by many increasingly accurate experimental test \cite{Bordag:2001qi,Bordag:2009zz} in the decades that followed Casimir's  paper, where he studied quantum fluctuations of the electromagnetic field in vacuum and discovered they produce an attraction of the plates.
Quantum fluctuations of other fields have also been found to produce Casimir forces, which are strongly dependent on the boundary conditions at the plates and on the geometry of the plates \cite{Boyer:1974,Boyer:1968uf}.
Dirichlet, Neumann or mixed boundary conditions are suitable for the scalar and electromagnetic fields and have been used extensively when investigating the Casimir effect due to fields of that kind, but cannot be used for fermionic fields \cite{Ambjorn:1981xv}. Bag boundary conditions, initially introduced to find a solution to confinement,  are used for fermionic fields \cite{Chodos:1974je,Johnson:1975zp} within the context of the Casimir effect.

The violation of Lorentz symmetry has become the subject of theoretical investigations in the last two decades, when several models that cause a space-time anisotropy have been proposed \cite{Horava:2009uw,Ferrari:2010dj,Ulion:2015kjx}, within the broader context of searching for a theory of quantum gravity, or string theory. While these models break Lorentz invariance at the Planck scale, repercussions of the space-time anisotropy can be observed at much lower energy through the Casimir effect, in the case of Lorentz symmetry breaking scalar fields \cite{Cruz:2017kfo,Cruz:2018bqt} and fermionic fields \cite{daSilva:2019iwn}, where the authors analyze the Casimir effect caused by a massless fermion field that violates Lorentz symmetry according to the Horava-Lifshitz model.

In recent years, authors have examined magnetic correction to the Casimir effect caused by charged scalar fields \cite{Cougo-Pinto:1998jun,Cougo-Pinto:1998fpo,Erdas:2013jga,Erdas:2013dha} and by Majorana fermion fields \cite{Erdas:2010mz} in Lorentz-symmetric space-time, and to the Casimir effect caused by a charged scalar field that breaks Lorentz invariance \cite{Erdas:2020ilo,Erdas:2021xvv}. However, a study of the magnetic effects on the Casimir effect caused by a charged fermion field that breaks Lorentz invariance has not been done. In this work I will investigate how a uniform magnetic field affects the Casimir effect due to a massless and charged fermion field
that violates Lorentz invariance according to the Horava-Lifshitz model. This fermion field will satisfy MIT bag boundary conditions on two identical parallel plates facing each other and the magnetic field will be uniform and perpendicular to the plates.

In Sec. \ref{2} of this paper I introduce the model of a fermion field that breaks Lorentz symmetry according to the theory of Horava-Lifshitz, discuss how to make the field satisfy bag boundary conditions on the plates and then obtain the field vacuum energy without a magnetic field and in the presence of the magnetic field. In Sec. \ref{3}, I use the zeta function regularization to calculate the Casimir energy of this system and the Casimir pressure it causes on the plates, in the presence of the external magnetic field. The Horava-Lifshitz theory associates a parameter with the Lorentz symmetry breaking mechanism, and this parameter is a positive integer called critical exponent. In this section two separate calculations are needed for even and odd values of the critical exponent, $\xi$, and they are presented in two subsections. I present a summary and discussion of my results in Sec. \ref{4}. In the Appendix I use the generalized zeta function technique to show how to calculate the Casimir energy for the special case where the critical exponent $\xi=1$.
%%%%%%%%%%%%%%%%%%%%%%%%%%%%%%%%%%%%%%%%%%%%%%%%%%%%%%%%%%%%%%%%%%%%%
\section{The model}
\label{2}

In this work I use $\hbar=c=1$ and study the Casimir effect due to a massless fermion field $\psi$ of charge $e$ that violates Lorentz symmetry according to the Horava-Lifshitz model, proposed by Horava in Ref. \cite{Horava:2009uw}. This theoretical model has a Lagrangian for the field $\psi$ that violates the Lorentz symmetry \cite{daSilva:2019iwn,Farias:2011aa} 
\begin{equation}
{\cal L}= {\bar \psi}\left[i\gamma^0\partial_t+i^{\xi}l^{\xi -1}(\gamma^i\partial_i)^\xi\right]\psi,
\label{L}
\end{equation}
where $\xi$, the critical exponent, is a positive integer and describes the violation of Lorentz symmetry, while the length parameter $l$ keeps $\cal L$ with the correct dimensionality. It is evident that, when $\xi = 1 $, the Lagrangian shown above is the standard Dirac Lagrangian for massless fermions and does not violate Lorentz symmetry. In Ref. \cite{daSilva:2019iwn} the authors obtain, from $\cal L$, the equation of motion of the fermion field $\psi$
\begin{equation}
\left[i\gamma^0\partial_t+i^{\xi}l^{\xi -1}(\gamma^i\partial_i)^\xi\right]\psi=0,
\label{DL}
\end{equation}
which can be viewed as a "Lorentz violating" Dirac equation when $\xi \ne 1$, and obtain all the solutions that obey MIT bag boundary conditions on two identical large parallel plates of area $L^2$, one located at $z=0$ and the other at 
$z=a$, with $L\gg a$. They find that any real value of the momentum components $k_x$ and $k_y$ is allowed, while only discrete values of $k_z$ are allowed. If the critical exponent $\xi$ is an odd number, then $k_z$ can only take the following discrete values
\begin{equation}
k_z=\left(n+\frac{1}{2}\right)\frac{\pi}{ a},
\label{kz_odd}
\end{equation}
where $n=0,1,2,\cdots$. If $\xi$ is even then $k_z$ can only take the following values
\begin{equation}
k_z={n\pi\over a},
\label{kz_even}
\end{equation}
where $n=1,2,3,\cdots$. 

The vacuum energy $E_0 = \langle 0 \vert H \vert 0 \rangle$ of the quantum field $\psi$ is obtained using the following Hamiltonian operator
\begin{equation}
H=i\int_Vd^3x \,\psi^\dagger \dot{\psi},
\label{H}
\end{equation}
where $V=L^2a$ is the volume contained between the two plates. It is straightforward to find
\begin{equation}
E_0=-C\left(L\over 2\pi\right)^2l^{\xi-1}\sum_n\int^\infty_{-\infty}dk_x \int^\infty_{-\infty}dk_y \left(k^2_x+k^2_y+k^2_z \right)^{\xi/2},
\label{E_0}
\end{equation}
where $C=2$ is the spin degeneracy factor and, when expressing $k_x$ and $k_y$ in polar coordinates, one immediately obtains
\begin{equation}
E_0=-{L^2\over \pi}l^{\xi-1}\sum_n \int^\infty_0dk \,k\left(k^2+k^2_z \right)^{\xi/2},
\label{E_02}
\end{equation}
where $k_z$ takes the form shown in Eq. (\ref{kz_odd}) if $\xi$ is odd, or takes the form shown in Eq. (\ref{kz_even}) if $\xi$ is even. The vacuum energy $E_0$ as shown in Eq. (\ref{E_02}) is divergent and, once it is renormalized 
through a regularization procedure, it is rendered finite and yields the Casimir energy $E_C$ of the fermionic quantum field confined between the two plates.

The novelty of this work is the presence of a uniform magnetic field $\vec B$ pointing in the $z$-direction, i.e. perpendicular to the plates. In the presence of a magnetic field the spin degeneracy is removed and, instead of the real-valued $k_x$ and $k_y$, discrete Landau levels appear. The vacuum energy is therefore 
\begin{equation}
E_0=E_++E_-, 
\label{E_03}
\end{equation}
where $E_+$ is the contribution of the spin up component of $\psi$ to the vacuum energy, and $E_-$ the contribution of the spin down component. I find
\begin{equation}
E_\pm=-{L^2 eB \over 2 \pi}l^{\xi-1}\sum_{n,\ell} \left[(2\ell+1)eB\pm eB+k^2_z \right]^{\xi/2},
\label{E_pm}
\end{equation}
where $(2\ell+1)eB$ are the Landau energy levels with $\ell = 0,1,2,\cdots$, and $\pm eB$ is the spin contribution.
%%%%%%%%%%%%%%%%%%%%%%%%%%%%%%%%%%%%%%%%%%%%%%%%%%%%%%%%%%%%%%%%%%%%%
\section{Casimir energy and pressure}
\label{3}

The Casimir energy $E_C$ is obtained by renormalizing the vacuum energy $E_0$. This renormalization process produces a finite Casimir energy and can be accomplished
using one of several regularization techniques. Ref. \cite{daSilva:2019iwn} used the Abel-Plana summation method, in this paper I use the zeta function technique.
Starting from Eqs. (\ref{E_03}) and (\ref{E_pm}), I define
\begin{equation}
E_C=-\lim_{\epsilon \to 0}{L^2 eB \over 2 \pi}l^{-(2s+1)}\sum_{n,\ell} \left\{\left[(2\ell+1)eB+ eB+k^2_z \right]^{-s}+\left[(2\ell+1)eB- eB+k^2_z \right]^{-s}\right\},
\label{E_C}
\end{equation}
where
\begin{equation}
s=-{\xi+\epsilon\over 2}.
\label{s}
\end{equation}
Using the following identity
\begin{equation}
z^{-s}={1\over \Gamma(s)}\int_0^\infty dt \,t^{s-1}e^{-zt},
\label{zgamma}
\end{equation}
I rewrite Eq. (\ref{E_C}) as
\begin{equation}
E_C=-\lim_{\epsilon \to 0}{L^2 eB \over 2 \pi}{l^{-(2s+1)}\over \Gamma(s)}\sum_{n} \int_0^\infty dt \,t^{s-1}e^{-k^2_zt} \sum_{\ell=0}^\infty \left[e^{-2(\ell+1)eBt}+ e^{-2\ell eBt} \right],
\label{E_C2}
\end{equation}
and, using this other identity
\begin{equation}
\sum_{\ell=0}^\infty \left[e^{-2(\ell+1)z}+ e^{-2\ell z} \right]=\coth (z),
\label{coth}
\end{equation}
I find
\begin{equation}
E_C=-\lim_{\epsilon \to 0}{L^2 eB \over 2 \pi}{l^{-(2s+1)}\over \Gamma(s)}\sum_{n} \int_0^\infty dt \,t^{s-1}e^{-k^2_zt}  \coth(eBt).
\label{E_C3}
\end{equation}
At this point, I will find analytical forms for the two asymptotic cases of weak magnetic field, $eB\ll a^{-2}$, and strong magnetic field, $eB\gg a^{-2}$. 

When the magnetic field is weak, I  use
\begin{equation}
\coth (eBt)\simeq {1\over eBt} + {eBt\over 3}-{(eBt)^3\over 45} + {\cal O}(B^5),
\label{coth2}
\end{equation}
into (\ref{E_C3}) and, after a change of integration variable, obtain
\begin{equation}
E_C\simeq-\lim_{\epsilon \to 0}{L^2 \over 2 \pi}{l^{-(2s+1)}\over \Gamma(s)} \left[ \Gamma (s-1)\sum_{n} (k^2_z)^{-(s-1)}+{e^2B^2\over 3}\Gamma (s+1)\sum_{n} (k^2_z)^{-(s+1)}
+{\cal O}(B^4)\right],
\label{E_C4}
\end{equation}
where, once the limit is taken, the first term will give the part of $E_C$ that does not depend on the magnetic field, while the second term, proportional to $B^2$, will give the leading magnetic correction to $E_C$.

In the case of strong magnetic field I use the following approximation
\begin{equation}
\coth (eBt)\simeq 1
\label{coth3}
\end{equation}
into (\ref{E_C3}) and, once I change the integration variable, I obtain
\begin{equation}
E_C\simeq-\lim_{\epsilon \to 0}{L^2 eB\over 2 \pi}l^{-(2s+1)}  \sum_{n} (k^2_z)^{-s}.
\label{E_C5}
\end{equation}

Since the discrete values of $k_z$ take a different form when the critical exponent is even, Eq. (\ref{kz_even}), or odd, Eq. (\ref{kz_odd}), I will calculate the two asymptotic limits of $E_C$ separately for each case.
%%%%%%%%%%%%%%%%%%%%%%%%%%%%%%%%%%%%%%%%%%%%%%%%%%%%%%%%%%%%%%%%%%%%%
\subsection{Casimir energy and pressure for even $\xi$}
\label{3_1}
When $k_z$ takes the form shown in Eq. (\ref{kz_even}), I evaluate the following sums in terms of the Riemann zeta function $\zeta_R(z)$
\begin{equation}
\sum_{n=1}^\infty (k^2_z)^{-(s-1)}=\left({\pi\over a}\right)^{-2(s-1)}\zeta_R(2s-2),
\label{kz1}
\end{equation}
\begin{equation}
\sum_{n=1}^\infty (k^2_z)^{-s}=\left({\pi\over a}\right)^{-2s}\zeta_R(2s),
\label{kz2}
\end{equation}
and
\begin{equation}
\sum_{n=1}^\infty (k^2_z)^{-(s+1)}=\left({\pi\over a}\right)^{-2(s+1)}\zeta_R(2s+2).
\label{kz3}
\end{equation}
Next, I use the sums evaluated in Eqs. (\ref{kz1}) and (\ref{kz3}), and insert these results into Eq. (\ref{E_C4}) to find
\begin{equation}
E_C\simeq-\lim_{\epsilon \to 0}{L^2 \over 2 \pi}{l^{-(2s+1)}\over \Gamma(s)} \left[ \Gamma (s-1)\left({\pi\over a}\right)^{-2(s-1)}\zeta_R(2s-2)+{e^2B^2\over 3}\Gamma (s+1)\left({\pi\over a}\right)^{-2(s+1)}\zeta_R(2s+2)
+{\cal O}(B^4)\right].
\label{E_Cw1}
\end{equation}
Last, I calculate the weak magnetic field limit of $E_C$ for $\xi = 2m$ with $m=1,2,\cdots$ by taking the limit for $\epsilon \to 0$ to obtain
\begin{equation}
E_C\simeq{L^2 \over 2 \pi}{l^{(2m-1)}} \left[ {1\over m+1}\left({\pi\over a}\right)^{2(m+1)}\zeta_R(-2m-2)+{e^2B^2\over 3}m\left({\pi\over a}\right)^{2(m-1)}\zeta_R(-2m+2)
+{\cal O}(B^4)\right],
\label{E_Cw2}
\end{equation}
and find that the part of $E_C$ that does not depend on the magnetic field vanishes for all even values of $\xi$, since $\zeta_R(-2n)=0$ for $n=1,2,\cdots$, thus confirming the result obtained in Ref. \cite{daSilva:2019iwn}. The magnetic correction, however, does not vanish for $\xi =2 $ ($m=1$) because $\zeta_R(0)=-{1\over 2}$ and, in this case, I find
\begin{equation}
E_C=-{L^2 l \over 12 \pi} e^2B^2,
\label{E_Cw3}
\end{equation}
where I use an equality since all higher order magnetic corrections vanish for $\xi = 2$. For even values of the critical exponent higher than $\xi =2$ the Casimir energy vanishes exactly, including low and high order magnetic corrections.

In the case of strong magnetic field, I insert into Eq. (\ref{E_C5}) the value of the sum I obtained in Eq. (\ref{kz2}) and find
\begin{equation}
E_C\simeq-\lim_{\epsilon \to 0}{L^2 eB\over 2 \pi}l^{-(2s+1)}  \left({\pi\over a}\right)^{-2s}\zeta_R(2s).
\label{E_Cs1}
\end{equation}
Once I take the limit $\epsilon \to 0$ and substitute $\xi =2m$, I obtain
\begin{equation}
E_C\simeq-{L^2 eB\over 2 \pi}l^{(2m-1)}  \left({\pi\over a}\right)^{2m}\zeta_R(-2m),
\label{E_Cs2}
\end{equation}
which vanishes for any $m\ge 1$.

The Casimir pressure is defined as
\begin{equation}
P_C=-{1\over L^2}{\partial E_C\over \partial a},
\label{P_C}
\end{equation}
and therefore, in the weak magnetic field limit, it vanishes for all even values of the critical exponent including $\xi =2$ since, in that case, the non-vanishing value of $E_C$ I obtained is independent of $a$. $P_C$ also vanishes in the strong magnetic field limit for all even values of $\xi$, since $E_C=0$ when the critical exponent is even.
%%%%%%%%%%%%%%%%%%%%%%%%%%%%%%%%%%%%%%%%%%%%%%%%%%%%%%%%%%%%%%%%%%%%%
\subsection{Casimir energy and pressure for odd $\xi$}
\label{3_2}
In this case $k_z$ takes the form shown in Eq. (\ref{kz_odd}), and I obtain the following results for the sums of Eqs. (\ref{E_C4}) and (\ref{E_C5})
\begin{equation}
\sum_{n=0}^\infty (k^2_z)^{-(s-1)}=\left({\pi\over a}\right)^{-2(s-1)}\left[2^{2(s-1)}-1\right]\zeta_R(2s-2),
\label{kz4}
\end{equation}
\begin{equation}
\sum_{n=0}^\infty (k^2_z)^{-s}=\left({\pi\over a}\right)^{-2s}\left(2^{2s}-1\right)\zeta_R(2s),
\label{kz5}
\end{equation}
and
\begin{equation}
\sum_{n=0}^\infty (k^2_z)^{-(s+1)}=\left({\pi\over a}\right)^{-2(s+1)}\left[2^{2(s+1)}-1\right]\zeta_R(2s+2).
\label{kz6}
\end{equation}
I use the values of two of these infinite sums into Eq. (\ref{E_C4}) to find, for weak magnetic field, the following
\begin{eqnarray}
E_C&\simeq&-\lim_{\epsilon \to 0}{L^2 \over 2 \pi}{l^{-(2s+1)}\over \Gamma(s)} \left[ \Gamma (s-1)\left({\pi\over a}\right)^{-2(s-1)}\left[2^{2(s-1)}-1\right]\zeta_R(2s-2)\right.
\nonumber \\
&&\left.+{e^2B^2\over 3}\Gamma (s+1)\left({\pi\over a}\right)^{-2(s+1)}\left[2^{2(s+1)}-1\right]\zeta_R(2s+2)
+{\cal O}(B^4)\right].
\label{E_Cw4}
\end{eqnarray}
To calculate the weak limit of $E_C$ for $\xi = 2m+1$, with $m=0,1,2,\cdots$, I take the limit for $\epsilon \to 0$ and find
\begin{equation}
E_C\simeq-L^2\!\left({ \pi l\over a} \right)^{2m}\!\!\left\{ {\pi^2\over (2m+3)a^3}\left[1-2^{-(2m+3)}\right]\zeta_R(-2m-3)+{2m+1\over 12\pi^2}e^2B^2a\left[1-2^{-(2m-1)}\right]\zeta_R(1-2m)
+{\cal O}(B^4)\!\right\}\! .
\label{E_Cw5}
\end{equation}
I now call $E^0_C$ the part of the Casimir energy that does not depend on the magnetic field, and $M_C$ the magnetic correction. 

For $\xi = 1 $ ($m=0$) the Lorentz symmetry is not violated by the fermion field and, since $\zeta_R(-3) = {1\over 120}$, I find 
\begin{equation}
E_C^0=-L^2{7\over 4}\left( {\pi^2\over 720 a^3}\right),
\label{E_C01}
\end{equation}
which agrees with Ref. \cite{daSilva:2019iwn} and with the earlier work on the Casimir energy of massless fermions \cite{DePaola:1999im}. The magnetic correction for $\xi =1$ appears to be divergent, since it is well known that $\zeta_R(1)=\infty$. However, this divergence is an artifact of a logarithmic dependence of $M_C$ on $a$. $M_C$, for $\xi =1$, can be calculated using another method, the generalized zeta function technique discovered by Hawking \cite{Hawking:1976ja}, to find a finite result \cite{Erdas:2010mz}. This calculation is shown in the Appendix, where I find
\begin{equation}
M_C=L^2 {e^2B^2a\over 12 \pi^2}\left[\gamma_E+\ln\left({2a\sqrt{eB}\over \pi }\right) \right],
\label{M_C}
\end{equation}
where $\gamma_E = 0.5772$ is the Euler-Mascheroni constant. $M_C$ is negative when $a\sqrt{eB}\ll 1$ (weak magnetic field).

When $\xi \ge 3$ ($m \ge1$), I find
\begin{equation}
E_C^0=-L^2{\pi^{(2m+2)}l^{2m}\over (2m+3) a^{(2m+3)}}\left[1-2^{-(2m+3)}\right]\zeta_R(-2m-3),
\label{E_C02}
\end{equation}
which is negative when $\xi=1,5,9,\cdots$, positive when $\xi=3,7,11,\cdots$, and fully agrees with what the authors of Ref.  \cite{daSilva:2019iwn}  obtained using a different regularization method.
The magnetic correction is
\begin{equation}
M_C\simeq-L^2 {(2m+1)\pi^{(2m-2)}l^{2m}\over 12a^{(2m-1)}}e^2B^2\left[1-2^{-(2m-1)}\right]\zeta_R(-2m+1)
+{\cal O}(B^4),
\label{M_C2}
\end{equation}
negative when $\xi=5,9,13\cdots$, positive when $\xi=3,7,11,\cdots$.

I use the definition of Casimir pressure of Eq. (\ref{P_C}) and, in the weak magnetic field limit, I find
\begin{equation}
P_C\simeq-{7\over 4}\left( {\pi^2\over 240 a^4}\right) - {e^2B^2\over 12 \pi^2}\left[1+\gamma_E+\ln\left({2a\sqrt{eB}\over \pi }\right) \right]+{\cal O}(B^4)
\label{P_C1}
\end{equation}
an attractive pressure for $\xi = 1$, and 
\begin{equation}
P_C\simeq-{\pi^{2}\over a^{4}}\left({\pi l\over a}\right)^{2m}\left[1-2^{-(2m+3)}\right]\zeta_R(-2m-3)-  {4m^2-1\over 12\pi^2}\left({\pi l\over a}\right)^{2m}e^2B^2\left[1-2^{-(2m-1)}\right]\zeta_R(-2m+1)
+{\cal O}(B^4),
\label{P_C2}
\end{equation}
for $\xi \ge 3$ ($m \ge1$). Notice that $\zeta_R(-2m-3)$ and $\zeta_R(-2m+1)$ are negative when $m$ is odd and are positive when $m$ is even, for $m\ge 1$,
thus the pressure is repulsive when $\xi =3, 7, 11, \cdots$, attractive when $\xi =5, 9, 13, \cdots$. Notice also that the magnetic correction increases the pressure in both cases of attractive and repulsive pressure. Finally, it is important to point out that, for any given value of $a\le \pi l$, the pressure is stronger for larger values of $\xi$.

In the limit of strong magnetic field I use the value of the sum of Eq. (\ref{kz5})  into Eq. (\ref{E_C5}) and, after taking the limit for $\epsilon \to 0$, obtain 
\begin{equation}
E_C\simeq{L^2 eB\over 2 }{(\pi l)^{2m}\over a^{(2m+1)}} \left[1-2^{-(2m+1)}\right]\zeta_R(-2m-1),
\label{E_Cs3}
\end{equation}
where I used $\xi =2m+1$ with $m$ being a non-negative integer. $E_C$ is negative when $\xi = 1,5,9,\cdots$, and it is positive when $\xi = 3,7,11,\cdots$. The strong magnetic field Casimir pressure is
\begin{equation}
P_C\simeq\left(m+{1\over 2}\right){ eB\over a^2 }\left({\pi l\over a}\right)^{2m} \left[1-2^{-(2m+1)}\right]\zeta_R(-2m-1),
\label{P_C3}
\end{equation}
and is attractive when $\xi = 1,5,9,\cdots$, while it is repulsive when $\xi = 3,7,11,\cdots$.
%%%%%%%%%%%%%%%%%%%%%%%%%%%%%%%%%%%%%%%%%%%%%%%%%%%%%
\section{Discussion and conclusions}
\label{4}

In this paper, I used the zeta function technique to investigate the Casimir effect of a Lorentz violating massless fermion field in the presence of a constant magnetic field perpendicular to the plates under the assumption that the quantum field satisfies bag boundary conditions at the plates. I obtained an integral expression for the Casimir energy (\ref{E_C3}) exact to all orders in the magnetic field $B$ and used it to find simple analytic expressions of the Casimir energy and pressure in the case of weak magnetic field, $eB\ll a^{-2}$, and strong magnetic field, $eB\gg a^{-2}$, for even values of the critical exponent, in Sec. \ref{3_1}, and for odd values of $\xi$, in Sec. \ref{3_2}.

I find that, when the critical exponent is even, the weak magnetic field correction $M_C$ to the Casimir energy is zero except for the case of $\xi = 2$, where $M_C$ takes a constant value, independent of the plate distance $a$. My study confirms that, for weak magnetic field and even values of the critical exponent, $E_C^0$ (the piece of the Casimir energy that does not depend on the magnetic field) vanishes as reported in Ref. \cite{daSilva:2019iwn}. I find that the strong magnetic field Casimir energy is zero for all even values of $\xi$. The Casimir pressure is also zero for all even values of $\xi$ and for weak and strong magnetic field.

When the critical exponent takes odd values I find that, for weak magnetic field, the values of  $E_C^0$ I report in Eqs. (\ref{E_C01}) and (\ref{E_C02}) agree with Ref. \cite{daSilva:2019iwn}, and I obtain their magnetic corrections
in Eqs. (\ref{M_C}) and (\ref{M_C2}). The Casimir pressure I obtain for weak magnetic field is attractive for $\xi=1,5,9,\cdots$, and repulsive for $\xi=3,7,11,\cdots$, as shown in Eqs. (\ref{P_C1}) and (\ref{P_C2}). Below are the weak magnetic field expressions of $E_C$ and $P_C$ for the first three odd values of $\xi$:
\begin{itemize}
  \item $\xi =1$
  \begin{equation}
E_C\simeq-L^2{7\over 4}\left( {\pi^2\over 720 a^3}\right)+L^2 {e^2B^2a\over 12 \pi^2}\left[\gamma_E+\ln\left({2a\sqrt{eB}\over \pi }\right) \right]+{\cal O}(B^4),
\label{D_1}
\end{equation}
\begin{equation}
P_C\simeq-{7\over 4}\left( {\pi^2\over 240 a^4}\right) - {e^2B^2\over 12 \pi^2}\left[1+\gamma_E+\ln\left({2a\sqrt{eB}\over \pi }\right) \right]+{\cal O}(B^4).
\label{D_2}
\end{equation}
 \item $\xi=3$
\begin{equation}
E_C\simeq L^2{\pi^{4}\left(1-2^{-5}\right)\over 1260 }{l^2\over a^5}+L^2 {l^{2}\over 96a}e^2B^2
+{\cal O}(B^4),
\label{D_3}
\end{equation}
\begin{equation}
P_C\simeq {\pi^{4}\left(1-2^{-5}\right)\over 256 }{l^2\over a^6}+ {l^{2}\over 96a^2}e^2B^2
+{\cal O}(B^4).
\label{D_4}
\end{equation}
 \item $\xi=5$
\begin{equation}
E_C\simeq -L^2{\pi^{6}\left(1-2^{-7}\right)\over 1680 }{l^4\over a^7}-L^2{\pi^{2}\left(1-2^{-3}\right)\over 288 }{l^4\over a^3}e^2B^2
+{\cal O}(B^4),
\label{D_5}
\end{equation}
\begin{equation}
P_C\simeq 
 -{\pi^{6}\left(1-2^{-7}\right)\over 240 }{l^4\over a^8}-{\pi^{2}\left(1-2^{-3}\right)\over 96 }{l^4\over a^4}e^2B^2+{\cal O}(B^4).
\label{D_6}
\end{equation}
\end{itemize}

When the critical exponent is odd and the magnetic field is strong the Casimir energy and pressure I obtain are reported in Eqs. (\ref{E_Cs3}) and (\ref{P_C3}) respectively.
Below are the strong magnetic field expressions of $E_C$ and $P_C$ for the first three odd values of $\xi$:
\begin{itemize}
  \item $\xi =1$
  \begin{equation}
E_C\simeq -L^2{eB\over 48 a},
\label{D_7}
\end{equation}
\begin{equation}
P_C\simeq -{eB\over 48 a^2}.
\label{D_8}
\end{equation}
 \item $\xi=3$
\begin{equation}
E_C\simeq L^2{\pi^{2}\left(1-2^{-3}\right)\over 240 }{l^2\over a^3}eB,
\label{D_9}
\end{equation}
\begin{equation}
P_C\simeq {\pi^{2}\left(1-2^{-3}\right)\over 80 }{l^2\over a^4}eB.
\label{D_10}
\end{equation}
 \item $\xi=5$
\begin{equation}
E_C\simeq -L^2{\pi^{4}\left(1-2^{-5}\right)\over 504 }{l^4\over a^5}eB,
\label{D_11}
\end{equation}
\begin{equation}
P_C\simeq -{5\pi^{4}\left(1-2^{-5}\right)\over 504 }{l^4\over a^6}eB.
\label{D_12}
\end{equation}
\end{itemize}
Notice how the numerical factor that determines the Casimir pressure increases as $\xi$ increases, from $1/48$ to $1/9.26$ to $1/1.07$.
%%%%%%%%%%%%%%%%%%%%%%%%%%%%%%%%%%%%%%%%%%%%%%%%%%%%%%%%%%%%%%%%%%%%%
%%%%%%%%%%%%%%%%%%%%%%%%%%%%%%%%%%%%%%%%%%%%%%%%%%%%%
\section{Appendix}
\label{5}
In this appendix I will show the details of the calculation of the Casimir energy in the presence of a weak magnetic field for the case of $\xi =1$, the Lorentz invariant case, using the generalized zeta function technique. The uniform magnetic field points in the $z$-direction, thus the generalized zeta function is constructed using the eigenvalues of the square of the following Dirac operators $D^{\pm}_E$ \cite{Erdas:2010mz}
\begin{equation}
D^{\pm}_E=-\partial^2_\tau+p^2_z - (\vec{p}-e\vec{A})^2_\perp\pm eB
\label{A1}
\end{equation}
where the subscript $E$ denotes the Euclidean time $\tau$, $\pm$ indicates spin up or down, $\vec A$ is the electromagnetic vector potential, and $\vec{p}_\perp =(p_x, p_y, 0)$. The eigenvalues of $D^+_E$ are
\begin{equation}
\left\{ k_0^2+{\pi^2\over a^2}\left(n+{1\over 2}\right)^2+2eB\left(\ell+{1\over 2}\right)+eB\right\},
\label{A2}
\end{equation}
where $k_0 \in \Re$, and $n$ and $\ell$ are non-negative integers. The eigenvalues of $D^-_E$ are the same, except that the last $+eB$ term is replaced by a $-eB$ term.
Using these eigenvalues I write the generalized zeta function as
\begin{eqnarray}
\zeta(s)&=&\mu^{2s}{L^2eB\over 4\pi^2}\sum_{n,\ell =0}^\infty\int_{-\infty}^\infty dk_0\left\{\left[  k_0^2+{\pi^2\over a^2}\left(n+{1\over 2}\right)^2+2eB\left(\ell+{1\over 2}\right)+eB\right]^{-s}\right.
\nonumber \\
&&\left.+\left[  k_0^2+{\pi^2\over a^2}\left(n+{1\over 2}\right)^2+2eB\left(\ell+{1\over 2}\right)-eB\right]^{-s}
\right\},
\label{A3}
\end{eqnarray}
where, as it is always done when using the generalized zeta function technique, the parameter $\mu$ with dimension of mass \cite{Hawking:1976ja} is introduced to 
keep $\zeta(s)$ dimensionless for all values of $s$. I use identities (\ref{zgamma}) and then (\ref{coth}) to write the generalized zeta function as
\begin{equation}
\zeta(s)={\mu^{2s}\over \Gamma(s)}{L^2eB\over 4\pi^2}\sum_{n =0}^\infty\int_{-\infty}^\infty dk_0\int_0^\infty dt \,t^{s-1}\exp\left[  -k_0^2 t-{\pi^2\over a^2}\left(n+{1\over 2}\right)^2t \right]\coth(eBt),
\label{A4}
\end{equation}
and do the $k_0$-integration to find
\begin{equation}
\zeta(s)={\mu^{2s}\over \Gamma(s)}{L^2eB\over 4\pi^{3/2}}\sum_{n =0}^\infty\int_0^\infty dt \,t^{(s-3/2)}\exp\left[-{\pi^2\over a^2}\left(n+{1\over 2}\right)^2t \right]\coth(eBt).
\label{A5}
\end{equation}
Next I use the weak field approximation of Eq. (\ref{coth2}) for $\coth (eBt)$, integrate over $t$ and obtain
\begin{equation}
\zeta(s)\simeq{\mu^{2s}\over \Gamma(s)}{L^2\over 4\pi^{3/2}}\left[\Gamma\left(s-{3\over 2}\right) \left({a\over \pi}\right)^{(2s-3)}\zeta_H\left(2s-3;{1\over 2}\right)+{e^2B^2\over 3} \Gamma\left(s+{1\over 2}\right) \left({a\over \pi}\right)^{(2s+1)}\zeta_H\left(2s+1;{1\over 2}\right)+{\cal O}(B^4)\right],
\label{A6}
\end{equation}
where $\zeta_H(s;z)$ is the Hurwitz zeta function of number theory, defined as
\begin{equation}
\zeta_H(s;z)=\sum_{n=0}^\infty (n+z)^{-s}.
\label{A7}
\end{equation}

The Casimir energy is obtained from the generalized zeta function in the following way \cite{Hawking:1976ja}
\begin{equation}
E_C=\zeta'(0),
\label{A8}
\end{equation}
therefore a power series expansion of $\zeta(s)$ valid for small $s$ will immediately produce $E_C$. I evaluate the following power series expansions
\begin{equation}
\left({\mu a\over \pi}\right)^{2s}{\Gamma\left(s-{3\over 2}\right)\over \Gamma(s)} \zeta_H\left(2s-3;{1\over 2}\right)\simeq-{7\sqrt{\pi}\over 720}s+{\cal O}(s^2),
\label{A9}
\end{equation}
\begin{equation}
\left({\mu a\over \pi}\right)^{2s}{\Gamma\left(s+{1\over 2}\right)\over \Gamma(s)} \zeta_H\left(2s+1;{1\over 2}\right)\simeq{\sqrt{\pi}\over 2}+\sqrt{\pi}\left[\gamma_E+\ln\left({2\mu a \over \pi}\right)
\right]s+{\cal O}(s^2),
\label{A10}
\end{equation}
and use them into (\ref{A6}) to evaluate the derivative of $\zeta(s)$ at $s=0$
\begin{equation}
\zeta'(0)\simeq-L^2{7\over 4}\left( {\pi^2\over 720 a^3}\right)+L^2 {e^2B^2a\over 12 \pi^2}\left[\gamma_E+\ln\left({2\mu a\over \pi }\right) \right]+{\cal O}(B^4).
\label{A11}
\end{equation}
The first term is the part of the Casimir energy independent of magnetic field and is the same I obtained, by a different method, in Eq. (\ref{E_C01}). The second term is the magnetic correction, which I call $M_C$ in Section \ref{3}. Notice that $M_C$ depends on the mass parameter $\mu$ which, in Refs.  \cite{Erdas:2021xvv,Cougo-Pinto:1998jun,Cougo-Pinto:1998fpo}, is taken to be $\mu=\sqrt{eB+M^2}$ where $M$ is the mass of the quantum field considered in those papers. In this paper the quantum field is massless, so I take $\mu=\sqrt{eB}$ and therefore
\begin{equation}
M_C\simeq L^2 {e^2B^2a\over 12 \pi^2}\left[\gamma_E+\ln\left({2 a\sqrt{eB}\over \pi }\right) \right]+{\cal O}(B^4).
\label{A12}
\end{equation}

%%%%%%%%%%%%%%%%%%%%%%%%%%%%%%%%%%%%%%%%%%%%%%%%%%%%%

%%%%%%%%%%%%%%%%%%%%%%%%%%%%%%%%%%%%%%%%%%%%%%%%%%%%%%%%%%%%%%%%%%%
\end{document}